\documentclass[superscriptaddress,secnumarabic,amssymb,amsmath,nobibnotes,aps,prd,showkeys,showpacs,twocolumn,nofootinbib]{revtex4-1}

\usepackage{orcidlink}
\usepackage{comment}

\usepackage{bm}
\usepackage{mathrsfs}
\usepackage[all]{xy}
\usepackage{epsfig,subfigure}
\usepackage{latexsym, amssymb, amsmath, amsfonts}
\PassOptionsToPackage{colorlinks=true,breaklinks=true}{hyperref}
\usepackage[english]{babel}
\usepackage[OT1]{fontenc}
\usepackage[utf8]{inputenc}
\usepackage{makeidx}
\usepackage[colorlinks=true,breaklinks=true]{hyperref}
\usepackage{graphicx, wrapfig, epsf}
\usepackage{float}

\usepackage[dvipsnames, table]{xcolor}
\hypersetup{urlcolor=BlueViolet,
	        citecolor=Plum,
	        linkcolor=OliveGreen}

\usepackage{natbib}
\usepackage{soul}

\hypersetup{
    colorlinks=true,
    linkcolor=Blue,
    filecolor=Blue,
    urlcolor=MidnightBlue,
    citecolor=Blue,
   pdftitle={Cooling age}
}

\newcommand{\dd}{ {\textrm d}}
\newcommand{\ee}{{\textrm e}}

\begin{document}

\title{The effects of strong gravity on the dispersion relation of massive particles\\ in the Kaluza\,--\,Klein theory}

\newcommand{\wigner}{HUN-REN Wigner Research Centre for Physics, 29-33 Konkoly-Thege Miklos Str., H-1121 Budapest, Hungary}

\author{Anna Horv\'ath\orcidlink{0000-0003-0328-1226}}
\thanks{Corresponding author}
\email[E-mail: ]{horvath.anna@wigner.hun-ren.hu}
\affiliation{\wigner}
\affiliation{Eötvös Loránd University, Pázmány Péter sétány 1/A, H-1117 Budapest, Hungary}

\author{Aneta Wojnar\orcidlink{0000-0002-1545-1483}}
\email[E-mail: ]{aneta.wojnar@uwr.edu.pl}
\affiliation{
Institute of Theoretical Physics, University of Wroc\l aw, pl. Maxa Borna 9, 50-206 Wroc\l aw, Poland}
\affiliation{Department of Theoretical Physics \& IPARCOS, Complutense University of Madrid, E-28040, 
Madrid, Spain 
}

\author{Gergely G\'abor Barnaf\"oldi\orcidlink{0000-0001-9223-6480}}
\email[E-mail: ]{barnafoldi.gergely@wigner.hun-ren.hu}
\affiliation{\wigner}

\begin{abstract}

We derive a modified dispersion relation for massive particles within the frameworks of five-dimensional Kaluza–Klein theory and general relativity, taking into account strong gravitational effects. The resulting effective mass depends on the curvature of the underlying phase space. Notably, in regions with strong gravitational fields, the effective mass may become imaginary, implying the possibility of particle decay induced by spacetime curvature.

\end{abstract}

\date{\today}

\maketitle

\section{Introduction}

\begin{figure*}[ht!]
     \includegraphics[scale=0.3]{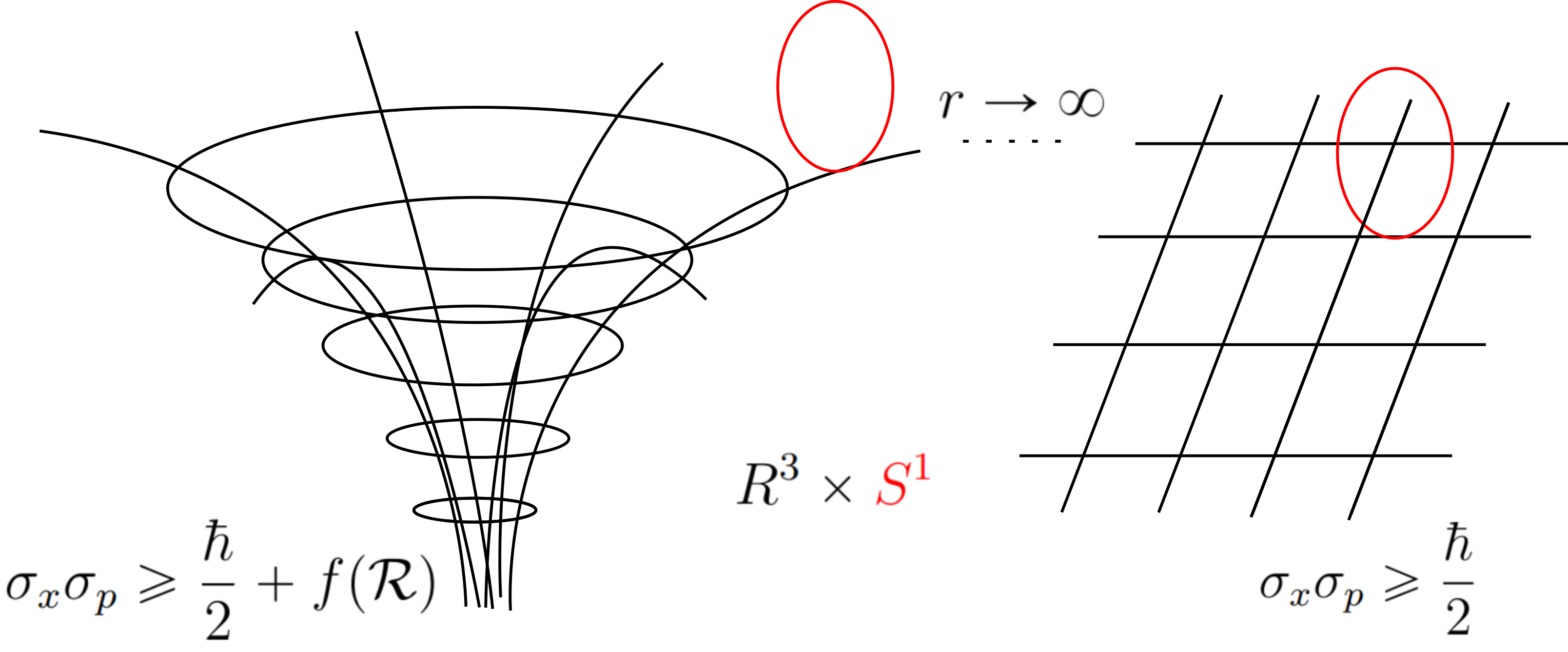}
     \caption{The effects of Kaluza\,--\,Klein theory and a strong gravitational field on the uncertainty relation of massive particles.}
     \label{fig:pic}
\end{figure*}

The interplay between gravity and the fundamental quantum interactions is a long researched topic in theoretical physics. Former has a geometrical description through Einstein's general relativity (GR), while the latter are well modelled by the standard model of particle physics. So far, experimental tests of these theories have found no discrepancies~\cite{LIGOScientific:2021sio,2019ApJ...875L...1E,PhysRevLett.130.071801,Erler_2019}, but there are many open questions which may open windows to beyond standard model theories~\cite{Laszlo_2018} and measurable effects~\cite{collaboration2025measurementpositivemuonanomalous, Krasznahorkay:2021joi}. Moreover at large scales, energies and in extreme circumstances their validity breaks down, giving rise to theories of quantum gravity~\cite{Ashtekar_2021}. 

A signature of fundamental theories of quantum gravity can be a modification to the Heisenberg uncertainty principle. Such a modification can be described from an effective point of view, without specifying the underlying theory. One class of modifications arises from the existence of a fundamental length or maximal momentum, called the generalized uncertainty principle~\cite{tawfik2015review}, which is in close relation to Born reciprocity~\cite{doi:10.1098/rspa.1938.0060}. The idea of a non-commutative spacetime~\cite{article} can also be connected to these type of modifications~\cite{Pachol:2023tqa}. In this study, our focus is on the so-called extended uncertainty principle, where the modifications appear due to the curvature of phase space~\cite{petruzziello2021gravitationally}. Within this formalism a quantum mechanical description of massive particles is possible with a modification to their dispersion relation.  

Besides taking into account quantum effects in a theory of gravity, our motivation was to study models, in which certain problems and puzzles of modern physics, such as the generally assumed existence of dark matter, galaxy rotation curves, or the nature of dark energy can be solved~\cite{Sahni:2004ai}. Our choice fell on extra-dimensional spacetime models~\cite{2021,csaki2004tasilecturesextradimensions}, which for example can reconcile the hierarchy problem~\cite{Randall_1999}. The simplest of them is the Kaluza\,--\,Klein (KK) theory which adding one extra compactified spatial dimension to the general relativistic spacetime~\cite{Overduin_1997}. Theodor Kaluza proposed it in 1921~\cite{1921SPAW.......966K}, unifying gravity with electromagnetism in a geometrical way. Since no sign of the fifth dimension has been detected so far, it is assumed that the metric components do not depend on the fifth coordinate. This rather unphysical condition was given an interpretation by Oskar Klein in 1926~\cite{1926ZPhy...37..895K}, suggesting that the extra dimension is microscopic in size, and curled up. Because of that, high energies are crucial to probe the effects of the fifth dimension. Furthermore, the periodic boundary condition on the circle introduces a $U(1)$ symmetry into the model~\cite{Horvath:2024raf,Horvath:2024qhk}, and leads to natural quantization.

On the other hand, the dimensional reduction of Kaluza\,--\,Klein theory introduces equations featuring a non-minimally coupled scalar field. Scalar-tensor theories, in various forms, have gained significant attention in recent years (see, e.g., \cite{faraoni2004scalar,kobayashi2019horndeski}), largely driven by the need to address cosmological phenomena that the standard concordance model, $\Lambda$CDM, struggles to fully explain \cite{abdalla2022cosmology,di2025cosmoverse}. The additional degree of freedom in these theories manifests on cosmological scales while being effectively suppressed in the Solar System (for a review of screening mechanisms, see e.g. \cite{hirano2019screening}). It has been subject to stringent constraints derived from observational data across various astrophysical contexts \cite{kobayashi2018relativistic} (for review, see \cite{olmo2020stellar}), since the screening mechanism is partially broken inside astrophysical bodies \cite{kobayashi2015breaking}.

These constraints include the detection of additional modes in gravitational wave signals \cite{mendes2018new}, scalarization effects in compact stars \cite{doneva2024spontaneous,vidal2025crystallized}, and investigations of the Sun \cite{saltas2019obtaining,saltas2022searching}, low-mass stars \cite{gomes2023early,chowdhury2019small,chowdhury2021modified}, and substellar objects such as brown dwarfs \cite{sakstein2015testing,leane2021exoplanets,kozak2023cooling}. Furthermore, Earth-based studies have also contributed to constraining these theories \cite{kozak2023earthquakes}. Simultaneously, black hole solutions in scalar-tensor theories have been extensively explored, leading to the prediction of "hairy" black holes \cite{brihaye2001dilatonic,doneva2022beyond,brihaye2022boson}.

The structure of the manuscript is as follows. In Section~\ref{sec:kk} we describe the Kaluza\,--\,Klein background, present the field equations, their solution, and provide a brief analysis on the geometry. Section~\ref{sec:adm_g} is dedicated to the quantum mechanical description of massive particles over a curved background provided by the solution of the KK theory. We present the Hamiltonian ADM formalism, and thoroughly analyze the phase space with respect to the parameter space. We present the extended uncertainty principle. In Section~\ref{sec:disp} the modified dispersion relation is calculated together with the effective mass of particles, which is compared to previous results. We conclude our work in Section~\ref{sec:conc}. 

\section{Kaluza\,--\,Klein theory in empty space}\label{sec:kk}

Kaluza\,--\,Klein theory can be written in the form of a five-dimensional, symmetric metric tensor, which has 15 independent components. These contain the 10 independent components of general relativity's $4\times 4$ symmetric metric tensor, and the vector potential of electromagnetism with 4 independent components. Thus, we are left with an extra degree of freedom, which can be associated with a scalar field. The five dimensional line element can be written as
\begin{equation}\label{5metric}
    \dd s^2 = \gamma_{\mu\nu}\dd x^\mu \dd x^\nu + 2\gamma_{\mu 5}\dd x^\mu \dd x^5 + \gamma_{55}\dd x^5\dd x^5\ ,
\end{equation}
where the indices, $\mu$ and $\nu$, run from 0 to 3, and the fifth direction is indicated by index '5' traditionally. It is usual to perform {\sl dimensional reduction} on the metric, by integrating out the fifth-dimensional coordinate, in order to get an effective, four-dimensional theory. Following the calculations presented in Coquereaux et al.~\cite{coquereaux1990theory}, one can rewrite the metric as
\begin{equation}
\label{eq:reduct}
    \dd s^2 = \underbrace{\left(\gamma_{\mu\nu}-\frac{\gamma_{\mu 5}\gamma_{\nu 5}}{\gamma_{55}}\right)}_{g_{\mu\nu}}\dd x^\mu \dd x^\nu + \underbrace{\gamma_{55} \vphantom{\frac{\gamma_{\mu 5}\gamma_{\nu 5}}{\gamma_{55}}}}_{{\ee^{2\sigma}}}\biggl(\dd x^5+\underbrace{\frac{\gamma_{\mu 5}}{\gamma_{55}}}_{\kappa A_\mu}\dd x^\mu\biggr)^2\ .
\end{equation}
The vector $A_\mu=A_\mu(x)$ is the electromagnetic vector potential, $\kappa$ is a constant, while $\sigma=\sigma(x)$ is the scalar field. Therefore, metric~\eqref{eq:reduct} can be written in terms of four-dimensional quantities as 
\begin{equation}
    \dd s^2 = g_{\mu\nu} \dd x^\mu \dd x^\nu + \ee^{2\sigma}\left(\dd x^5+\kappa A_\mu \dd x^\mu\right)^2\ .
\end{equation}
The five-dimensional model is simply a geometrical generalization of Einstein's theory, while the effective, four-dimensional field equations are modified by functions of the quantities specified by the braces of Eq.~\eqref{eq:reduct}. One implements the  dimensional reduction, on one hand, because measurements are performed in the usual four dimensions that one experiences\footnote{That is, a 4 dimensional case decomposed to a 3+1 representation after choosing an observer indicating an arrow of time.}, on the other hand, since it makes possible to compare the theory to the four-dimensional formalism by e.g. introducing a fifth force~\cite{Fischbach:1985tk,Fischbach:1992fa}. 

The action of the  theory in the Jordan conformal frame is given by
\begin{equation}\label{action1}
    S= \int \frac{\sqrt{-g}}{16\pi \mathbb{G}\ee^{-\sigma}}\left[
    R-\frac{\kappa^2}{4}\ee^{2\sigma}F_{\mu\nu}
 F^{\mu\nu}-2\ee^{-\sigma}\Box \ee^{\sigma} 
    \right]\dd^4x\ ,
\end{equation}
where the third term involving $\Box \ee^\sigma=\nabla^\mu\partial_\mu\ee^\sigma$ is a total divergence, therefore it can be removed. $\mathbb{G}$ is the 5-dimensional gravitational constant while the 4-dimensional one is defined as $\mathcal{G}=\mathbb{G}\ee^{-\sigma}$. Note that $\mathcal{G}$ is a variable if $\sigma$ is not a constant. This action provides a gravitational model in which gravity is coupled to electromagnetism
and a scalar field represented by $\sigma$.

However, it is more convenient to rewrite action~\eqref{action1} by performing a conformal transformation of the form
\begin{equation}
    \tilde{g}_{\mu\nu}(x) = \ee^{\chi} g_{\mu\nu}(x)\ , 
\end{equation}
where $\tilde g$ is the a conformal metric, and $\chi=\chi(x)$ is a scalar function. By setting $\chi = \sigma$, the variability of the four-dimensional gravitational constant can be transformed out of the action: $\mathcal{G}|_{\chi = \sigma}=\mathbb{G}e^{\chi-\sigma}=\mathbb{G}$. Thus, one gets the action in the Einstein frame\footnote{Note, quantities in the Einstein frame are denoted by '$\sim$'} form
\begin{equation}\label{action2}
\begin{aligned}
     S= \int \frac{\sqrt{-\tilde{g}}}{16\pi \mathbb{G}}&\left[ 
    \tilde{R}-\frac{3}{2}\partial_\mu\sigma\tilde{\partial}^\mu\sigma\right. \\
   &-\left.   \frac{\kappa^2}{4}e^{3\sigma}F_{\mu\nu}
 \tilde{F}^{\mu\nu}+ \tilde{\nabla}_\mu \tilde{\partial}^\mu\sigma 
    \right]\dd^4x\ ,
\end{aligned}
\end{equation}
which resembles general relativistic equations with an electromagnetic and scalar field source. The $\tilde{\nabla}$ is the covariant derivative with respect to the metric $\tilde{g}_{\mu\nu}$, and $\tilde{\partial}^\mu=\tilde{g}^{\mu\nu}\partial_\nu$. Notice that the frames are not equivalent physically in general~\cite{Dabrowski:2008kx,Fujii:2003pa}. 
 Moreover, we have not included arbitrary matter fields here (apart from the electromagnetic field, which is conformally invariant, and the scalar field itself). Note that the arbitrary matter would acquire a coupling under conformal transformation.

From now on, let us consider a static, spherically symmetric empty spacetime, in which 
the five-dimensional energy-momentum tensor is zero. However, let us notice that the effective four-dimensional one, in general, can contain contributions from both the vector potential, $A_\mu$, and the scalar field, $\sigma$. Here, we restrict ourselves to the electromagnetic free theory: $A_\mu=0$, for which the field equations read \cite{coquereaux1990theory}:
\begin{equation}
\begin{aligned}
\label{eq:fe}
    \tilde{R}_{\mu\nu} &= 8\pi \mathbb{G}\left(\tilde{T}_{\mu\nu}-\frac{1}{2}\tilde{g}_{\mu\nu}\tilde{T}^{\rho}_\rho \right) \\
    \tilde{\nabla}_\mu\tilde{\partial}^\mu\sigma&= 0\ ,
\end{aligned}
\end{equation}
where
\begin{equation}
    \tilde{T}_{\mu\nu} = \frac{3}{16\pi \mathbb{G}}\left(\partial_\mu\sigma\partial_\nu\sigma - \frac{1}{2}\tilde{g}_{\mu\nu}\partial_\rho\sigma\tilde{\partial}^\rho\sigma\right).
\end{equation}
 For a spherically symmetric, static spacetime, the effective four-dimensional metric can be written in a general, diagonal form:
\begin{equation}
    \begin{aligned}
    \dd s^2 =&- \ee^\nu \dd t^2 + \ee^{-\nu}\dd r^2+\ee^{\lambda-\nu}\dd \Omega^2\ \\
    & \hspace{1truecm} \text{with} \quad \dd \Omega^2=\dd\theta^2+\sin^2{\theta}\dd\phi^2\ ,
\end{aligned}
\end{equation}
where $\nu=\nu(r)$ and $\lambda=\lambda(r)$. The exact solution of this form is the generalized Schwarzschild metric \cite{coquereaux1990theory}:
\begin{equation}\label{eq:metric_ef}
\begin{aligned}
    \dd s^2 = &-\left(1-\frac{a}{r}\right)^{\frac{b}{a}}\dd t^2 +
    \left(1-\frac{a}{r}\right)^{-\frac{b}{a}}\dd r^2 \\ 
    &+r^2 \left(1-\frac{a}{r}\right)^{1-\frac{b}{a}}\dd\Omega^2\ ,
\end{aligned}
\end{equation}
where $a$, $b$ and $d$ are constants satisfying
\begin{equation}\label{eq:constraint}
    a^2=b^2+3d^2\ ,
\end{equation}
being a result of the field equations~\eqref{eq:fe}. The scalar field can be expressed as 
\begin{equation}\label{eq:sigma}
\begin{aligned}
    \sigma(r) &= \frac{d}{a}\ln \left(1-\frac{a}{r}\right) \\
    d &= \sigma(r)'(r^2-ar) = \text{const}\ ,
\end{aligned}
\end{equation}
where the prime ($'$) denotes the derivative with respect to the radial coordinate $r$.

 Notice that the Einstein frame metric~\eqref{eq:metric_ef} does not have any explicit dependence on the parameter $d$, but is constrained by Eq.~\eqref{eq:constraint}.
The Ricci scalar curvature of the conformal spacetime has the form
\begin{equation}\label{eq:sptr_ef}
    \tilde{R} = \frac{a^2-b^2}{2r^4}\left(1-\frac{a}{r}\right)^{\frac{b}{a}-2}\ .
\end{equation}
On the other hand, it is straightforward to transform the obtained result, metric~\eqref{eq:metric_ef}, back to the Jordan frame:
\begin{equation}\label{metric}
\begin{aligned}
    \dd s^2 = &-\left(1-\frac{a}{r}\right)^{\frac{b-d}{a}}\dd t^2 +
    \left(1-\frac{a}{r}\right)^{-\frac{b+d}{a}}\dd r^2\\ &+ r^2 \left(1-\frac{a}{r}\right)^{1-\frac{b+d}{a}}\dd\Omega^2\ .
\end{aligned}
\end{equation}
For this metric the Ricci scalar spacetime curvature is 
\begin{equation}\label{eq:sptr}
    R = \frac{a^2-b^2-3d^2}{2r^4}\left(1-\frac{a}{r}\right)^{\frac{b+d}{a}-2}\ ,
\end{equation}
which equals zero by Eq.~\eqref{eq:constraint}, as expected for the empty space-time. Note that the spacetime curvature is non-zero for the Kaluza\,--\,Klein case in the Einstein frame, Eq.~\eqref{eq:sptr_ef}, except for the usual Schwarzschild solution, where $a=b$ and $d=0$. This, again, shows the qualitative difference of the two frames.

\subsection{Properties of the Kaluza\,--\,Klein spacetime}
\label{sec:adm}

The Kaluza\,--\,Klein effective four-dimensional metric significantly differs in structure from the usual general relativistic case. The Schwarzschild solution is recovered from this theory, when $a=b=2M$, with $M$ being the mass parameter of the central object in geometrized units ($c=1$, $\mathcal{G}|_{\chi = \sigma}=\mathbb{G}=1$). The question whether this modifies gravity on the Newtonian scale naturally arises. 

\begin{figure}[ht]
     \includegraphics[scale=0.55]{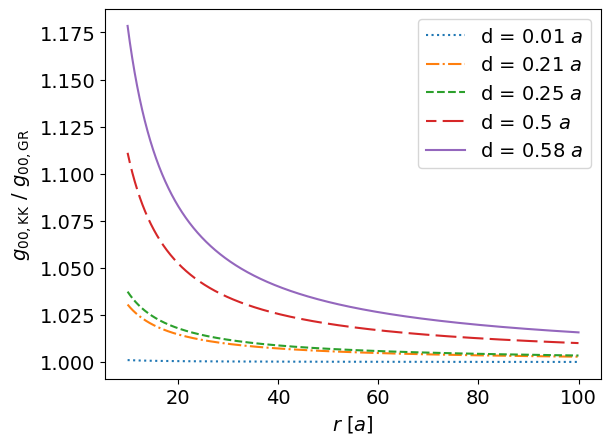}
     \caption{The ratio of the $g_{00}$ terms of the KK and the GR Schwarzschild solutions' metric in the Jordan frame as a function of the radial coordinate $r$, in units of $a$.}
     \label{fig:ratio}
\end{figure}

If one wishes to keep the scalar field and the constants of integration real (see Fig.~\ref{fig:param_space} showing the full parameter space in Appendix~\ref{appendix_new1}), the allowed values of the parameter $d$ are $d\in (-\sqrt{a^2/3},\sqrt{a^2/3})$. Also, the value of $b$ is taken to be positive here, since this case is what resembles the usual Schwarzschild solution. Plotting the ratio of the $g_{00}$ terms of the KK and the GR Schwarzschild solutions, $\left(1-\frac{a}{r}\right)^{\frac{b-d}{a}-1}$ in Figure~\ref{fig:ratio}, one can see that the correction is less than three percent far away (for these $d$ at $r\approx100a$) from the center, however, in the vicinity of the object ($r\lesssim$ 30a), it becomes significant. Note that in Fig.~\ref{fig:ratio} the special values of $d$, $3/14a\approx 0.21a$ and $a/\sqrt{3}$ is slightly larger than $0.58a$. The GR case, $d=0$, is indistinguishable from the technically calculable $d=0.01a$. 

To analyze the structure of spacetime in more detail, let us consider a normalized ($u^\mu u_\mu = -1$), non-rotating vector field $u^\alpha$. Then, $u^\alpha$ specifies the movement of an observer admitting a foliation of spacetime according to the Arnowitt\,--\,Deser\,--\,Misner (ADM) formalism (see, e.g. Refs.~\cite{Gourgoulhon:2007ue,Borowiec:2013kgx}). This means that we can locally decompose the spacetime metric in such a way that the line element of metric~\eqref{metric} can be expressed as
\begin{equation}\label{split}
\begin{aligned}
   \dd s^2=&-N^2\left(\dd x^0\right)^2 \\
   &+h_{ij}(N^i\dd x^0+\dd x^i)(N^j\dd x^0+\dd x^j)\ ,
\end{aligned}
\end{equation}
where $N$ is the lapse function, $N^i$ is the shift vector and $h_{ij}$ is the induced metric on 3-dimensional hypersurfaces of constant $x^0$, with $i,j$ running from 1 to 3. In our specific case, $N=\left(1-\frac{a}{r}\right)^{\frac{b-d}{2a}}$, 
$N^i=0$, and $h_{ij}$ is simply the spatial part of $g_{\mu\nu}$. With this formalism, we can now examine the ADM mass. It provides the total energy contained in each 3-dimensional leaf  $\Sigma_t$ of a foliation of spacetime $(\Sigma_t)_{t\in \mathbb{R}}$, defined as 
\begin{equation}\label{adm_gen}
    M_\mathrm{ADM} := -\frac{1}{8\pi} \lim_{S_\mathrm{t} \to \infty} \oint_{S_\mathrm{t}} (k-k_0) \sqrt{q} \dd^2x\ ,
\end{equation}
where $S_\mathrm{t}$ is a 2-dimensional surface at infinity, $k$ is the trace of the extrinsic curvature of $S_\mathrm{t}$ embedded in the curved metric containing the mass and energy to be evaluated, $k_0$ is the trace of the extrinsic curvature of $S_\mathrm{t}$ embedded in the asymptotic flat metric and $\sqrt{q} \dd^2x$ is the surface element~\cite{Gourgoulhon:2007ue}. 
A straightforward calculation provides
\begin{equation}\label{eq:adm_int}
    M_\mathrm{ADM}=\frac{1}{16\pi} \lim_{r \to \infty} \oint_{r = \text{const}} B(r)r^2\sin{\theta}\dd\theta \dd\phi\ ,
\end{equation}
where 
\begin{equation}\label{eq:adm_integrand}
    B(r)= 2\frac{b+d}{r^2}\left( 1-\frac{a}{r}\right)^{-\frac{b+d}{a}}\ .
\end{equation}
Rewriting Eq.~\eqref{eq:adm_integrand} as a function of $d$ based on Eq.~\eqref{eq:constraint} and assuming $b>0$, one arrives at 
\begin{equation} \label{eq:adm}
\begin{aligned}
 B(r)&=  \frac{2}{r^2} (d+\sqrt{a^2-3d^2})\left( 1-\frac{a}{r}\right)^{-\frac{d+\sqrt{a^2-3d^2}}{a}} \\ 
    &\approx \frac{2a}{r^2}\left(1+\frac{d}{a}\right)-\frac{3a}{r^2}\left(\frac{d}{a}\right)^2\ .& 
\end{aligned}
\end{equation}
In the above equation in the second line we have taken the weak field limit up to the second order in $d$ and $r$, that is, $r\to\infty$ and $d\to0$, using power series expansion. In case of the usual Schwarzschild metric, where $a=2M$ and $d=0$, Eq.~\eqref{eq:adm} reduces to $4M/r^2$, which gives the Newtonian mass of the central object upon integration. Therefore, we assume that $a$ is positive. In the Kaluza\,--\,Klein case, we also have a contribution from the scalar field $\sigma$, whose presence in the considered solution is related to the constant model parameter $d$. Taking into account only the first order term, the mass, observable far away from the central object is given by
\begin{equation}\label{adm_mass}
    M_\mathrm{ADM}\approx\frac{a}{2}\left(1+\frac{d}{a}\right)\ .  
\end{equation}

\section{Non-relativistic massive particle on a curved background} \label{sec:adm_g}

To analyze the motion of a free test particle on a curved, effectively 4-dimensional background, we recall the procedure provided in Petruzziello et al.~\cite{petruzziello2021gravitationally}. The Lagrangian describing the effective dynamics of a massive particle on the curved four-dimensional background is given by
\begin{equation}
L=-m\sqrt{-g_{\mu\nu}(x)\dot{x}^\mu \dot{x}^\nu}\ ,\label{lagrangian2}
\end{equation}
where $m$ is the mass of the test particle, $g_{\mu\nu}$ is the background metric, while the four-velocity of the particle is given by $u^\mu=\dot{x}^\mu=\dd x^\mu/\dd\tau$, where $\tau$ denotes the proper time along the curve. Assuming that the particle is not rotating, we can locally slice the spacetime~\eqref{metric} according to the ADM procedure. 
This formalism allows for a quantum mechanical treatment, since time evolution becomes possible along the separate $x^0$ coordinate specifying curved, 3-dimensional embeddings into 4 dimensions. 

We can rewrite the above Lagrangian in the ADM formalism as
\begin{equation}
    L=-m\sqrt{N^2\left(\dot{x}^0\right)^2-h_{ij}\left(N^i\dot{x}^0+\dot{x}^i\right)\left(N^j\dot{x}^0+\dot{x}^j\right)}\ ,
\end{equation}
where $i,j\in \{1,2,3\}$. Considering only the non-relativistic part of the above Lagrangian  
provides
\begin{equation}\label{eq:lnr}
    L_\mathrm{NR}=\frac{m}{2}\left(N^i+\dot{x}^i\right)\left(N^j+\dot{x}^j\right)G_{ij}-mN\ ,
\end{equation}
where one defines $G_{ij}:=h_{ij}/N$, and fixes the gauge to $x^0=\tau$, specifying a co-moving observer. It can be shown that the above Lagrangian is invariant under gauge transformations, while the canonical momentum,
\begin{equation}
    \pi_i=\frac{\partial}{\partial \dot{x}^i}L_\mathrm{NR}=mG_{ij}\left(\dot{x}^j+N^j\right)\ 
\end{equation}
is not, that is, it does not provide observables. Thus, let us define the gauge invariant physical momentum as
\begin{equation}
    p_i:=\pi_i-mN^jG_{ij}\label{phys}\ ,
\end{equation}
providing the Hamiltonian 
\begin{equation}
    H_{NR}=\frac{1}{2m}p_ip_jG^{ij}+m\left(N-\frac{N^iN^j}{2}G_{ij}\right).\label{effhamiltonian}
\end{equation}

For the quantum mechanical description~\cite{petruzziello2021gravitationally} the Hilbert space measure defining the scalar product, $\langle \psi | \phi \rangle = \int \dd \mu \psi^*\phi$ on a curved background can be written as
\begin{equation}\label{measure}
\dd\mu=\sqrt{G}\dd^3x\ ,
\end{equation}
where $G=\mathrm{det }G_{ij}$. The physical momentum operator takes the form
\begin{equation}
\begin{aligned}
    \hat{p}_i\psi =& (\hat\pi_i-m\hat{N}^j\hat{G}_{ij})\psi
    \\
    \equiv&-\left[i\hbar\left(\partial_i+\frac{1}{2}\Gamma^j_{ij}(G)\right)+mN^jG_{ij}\right]\psi\ ,
    \label{momentum}
\end{aligned}
\end{equation}
where $\hat\pi_i$ is defined by the second equality in Eq.~\eqref{momentum}, and $\Gamma^i_{jl}(G)$ is the Levi-Civita connection with respect to the effective metric $G_{ij}$. Perturbing the metric, the measure, and the physical momentum, together with its covariant form and square provides the uncertainty relation of a massive non-relativistic particle on a curved 4-dimensional background  
\begin{align}
    \sigma_\mathrm{p}\rho\gtrsim &\pi\hbar\left[1-\frac{\rho^2\mathcal{R}|_{x_0}}{12\pi^2}+ \left.\xi\frac{\rho^4}{\lambda_C^2}\bar{\nabla}_jN_i\bar{\nabla}^jN^i \right|_{x_0}\right]\ .\label{finalunc}
\end{align}
In the above, $\sigma_\mathrm{p}=\sqrt{\langle \hat{p}^2\rangle-\langle \hat{p}_i\rangle\langle \hat{p}^i\rangle}$ is the momentum uncertainty, while $\rho$ is the radius of the 3-dimensional geodesic ball, in which the quantum states are confined, and $x_0$ it its center\footnote{Note that $\rho$ should rather be interpreted as an uncertainty and it does not represent the standard deviation of position, see \cite{dabrowski2020asymptotic}.}. The Ricci scalar $\mathcal R$ of the phase space is derived from the metric $G_{ij}$ evaluated at $x_0$. The covariant derivative, $\bar{\nabla}_i$, is evaluated with respect to the canonical connection of $G_{ij}$. The constant $\xi$ equals $(2-3/\pi^2)/9$, while $\lambda_\mathrm{C}=2\pi \hbar/m$ is the Compton wavelength
~\cite{dabrowski2020asymptotic,petruzziello2021gravitationally,wagner2022relativistic}. 

Notice that for the considered Kaluza\,--\,Klein spacetime with a co-moving observer, the shift vector $N^i$ vanishes. Moreover, the non-relativistic, relativistic and ultra-relativistic cases will be the same, since the ultrarelativistic corrections to the uncertainty relation depend on $N^i$ \cite{wagner2022relativistic}, and in its absence, Eq.~\eqref{finalunc} reduces to the non-relativistic case. Due to this fact, our further results can also be applied to relativistic massive particles for the chosen observer. 
\begin{figure*}[ht!]
     \includegraphics[scale=0.5]{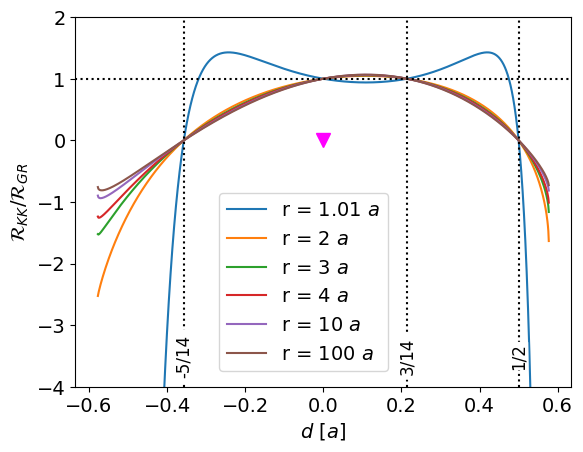}
     \includegraphics[scale=0.5]{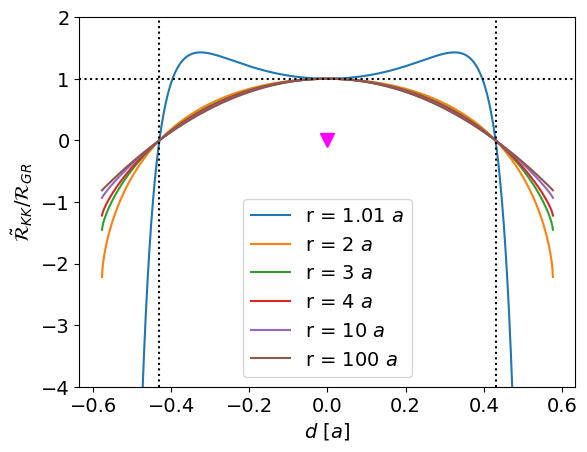}
     \caption{Phase space '{\sl cat}'. Left: Jordan frame, right: Einstein frame. Phase space curvature ratio of the Kaluza\,--\,Klein theory and usual general relativity. Different colors correspond to different $r$ radial coordinate values, while the vertical dashed lines denote special values of $d$ (at $d=\pm \sqrt{(a-4/9a)/3}$ for the Einstein '{\sl cat}'). The '{\sl nose}' indicates the (0,0) coordinate.}
     \label{fig:ph_ratio2}
\end{figure*}

\subsection{The Kaluza-Klein phase space}

The Kaluza\,--\,Klein spacetime in the Jordan frame is given by Eq.~\eqref{metric}.  
It is now straightforward to obtain the Ricci scalar curvature $\mathcal{R}$ of the phase space ---which is derived from $G_{ij}$, defined in Eq.~\eqref{eq:lnr}:
\begin{equation}\label{curv2}
    \mathcal R = \frac{4a^2-(3b+d)^2}{8r^4}\left(1-\frac{a}{r}\right)^{\frac{3b+d}{2a}-2}.
\end{equation}
On the other hand, the Einstein frame case provides
(calculated from $\tilde{G}_{ij}$)
\begin{equation}\label{curv3}
    \tilde{\mathcal R} = \frac{4a^2-9b^2}{8r^4}\left(1-\frac{a}{r}\right)^{\frac{3b}{2a}-2}.
\end{equation}
Both phase space curvatures are non-zero. Note that none of them vanishes even for the usual general relativistic case, $a=b$ ($d=0$) providing the general relativistic Schwarzschild expression
\begin{equation}\label{curv4}
    \mathcal R_\mathrm{GR}=\mathcal R \equiv   \tilde{\mathcal R} = -\frac{5 a^2}{8r^4}\left(1-\frac{a}{r}\right)^{-\frac{1}{2}}\ .
\end{equation}

The ratio of the Kaluza\,--\,Klein and the general relativistic phase space curvatures as a function of the parameter $d$ for $b>0$ is plotted in Figure~\ref{fig:ph_ratio2} for both the Jordan (left) and the Einstein (right) frames. We consider different values of the radial coordinate $r$, adjusted to a given frame.{
Since the phase space curvatures are nonzero, one can see that both theories generate gravitational corrections to the uncertainty relation and momenta, even when the shift vectors are zero.

In the Schwarzschild limit, see Eq.~\eqref{curv4}, the curvature is always negative, considering the region where $a>r$. However, in the Kaluza\,--\,Klein case the sign changes at the outer base of the '{\sl ears}' of both '{\sl cats}', 
making the '{\sl whiskers}' region significantly different from the general relativistic case. 

The symmetry around $d=0$ of the Einstein frame '{\sl cat}' in Fig.~\ref{fig:ph_ratio2} is inherited from the fact that  $\tilde{\mathcal{R}}$ does not depend on the sign of $d$. Moreover, unlike in the Jordan frame, we have only one equivalence to general relativity for all $r$, at the top of the '{\sl head}', where $d=0$. In the Jordan frame, one deals with an additional point at $d=3/14$, for which both curvatures coincide at all radii. Note that in both frames, there exist two additional non-zero values of $d$ for which the phase space curvatures coincide.  
Close to the horizon, at around $r\approx1.08a$ this degeneracy expands to a larger span of $d$-values --- the '{\sl ears}' of the '{\sl cat}' flatten out.

\begin{figure}[t]\centering
     \includegraphics[scale=0.55]{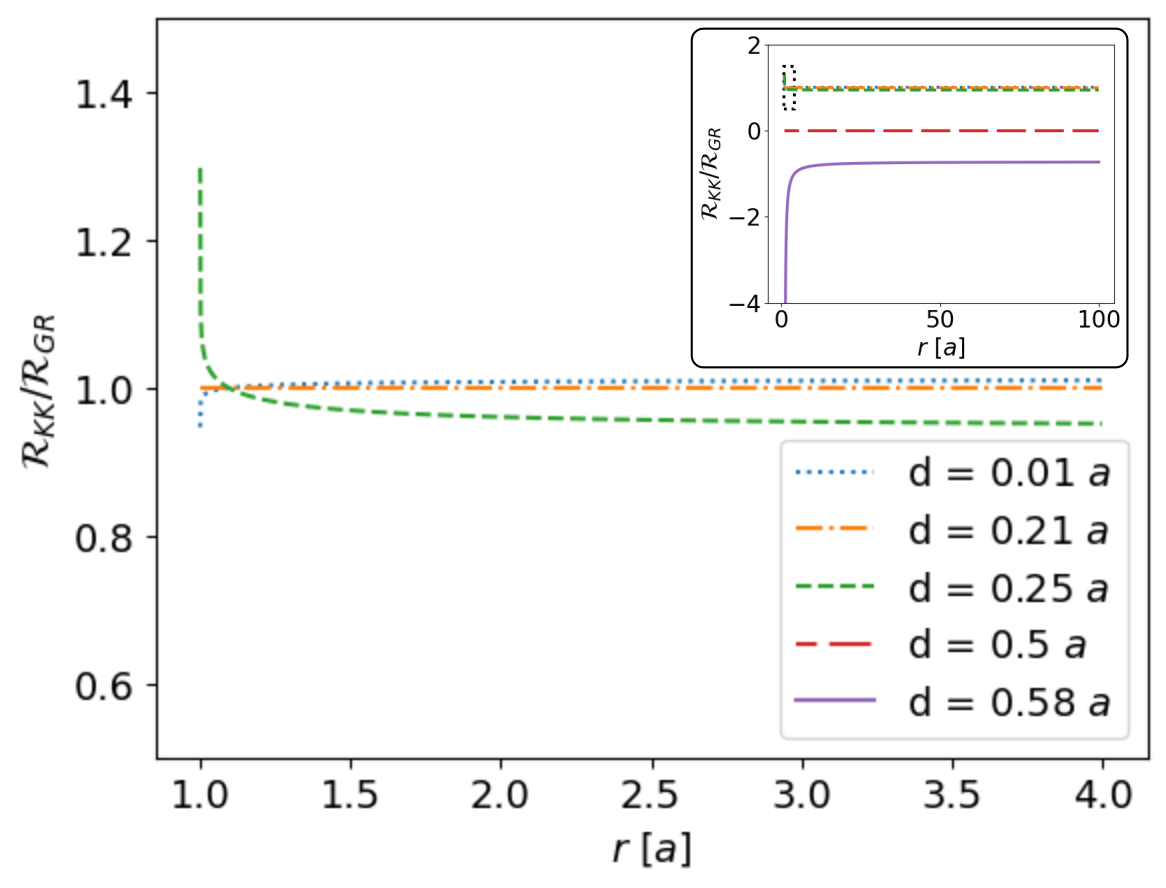}
     \caption{Phase space curvature ratio of the Kaluza\,--\,Klein and the general relativistic Schwarzschild metric calculated from $G_\mathrm{ij}$ for different values of parameter $d$ (connected to the scalar field) as a function of the radial coordinate $r$, given in the Jordan frame. The larger plot is a magnified part of the smaller one indicated by a dotted rectangle.}
     \label{fig:ph_ratio1}
\end{figure}

The ratio of the Kaluza\,--\,Klein and the general relativistic phase space curvatures (in the Jordan frame) as a function of the radial coordinate $r$ is plotted in Figure~\ref{fig:ph_ratio1} for discrete $d$-values ($b>0$). The different curves saturate to separate constant values far away from the centre, however, at small $r$ values they diverge, except for special cases. Note, the special values of $d$, $3/14a\approx 0.21a$ and $a/\sqrt{3}$ are slightly larger than $0.58a$. The GR case, $d=0$, is indistinguishable fromfrom $d=0.01a$.

The saturation to different constant values shows how both curvatures decrease with approximately the same power of $r$ asymptotically (compare Eq.~\eqref{curv2} and Eq.~\eqref{curv4}), but are different quantitatively. This might allow for constraining the Kaluza\,--\,Klein theory even further away from the gravitating object, in the weak-field limit.

\subsection{The Kaluza\,--\,Klein parameter space}

Let us briefly investigate the behaviour of the theory in different regions of the parameter space. In usual Schwarzschild coordinates, parameter $a$ is associated with the black hole mass. However, it does not have to be so in the Kaluza\,--\,Klein case. We use it as a unit in our analysis, and treat $b$ and $d$ as the model parameters. Their connection is set by Eq.~\eqref{eq:constraint}, and $d$ defines $b$ up to its sign, which we set to be positive. The positive value keeps the metric coefficients close to the general relativistic one, particularly when the values of $d$ are small.

A didactical sketch of the different regimes of the phase space curvature in the Jordan frame as a function of $d$, measured in units of $a$ is shown in Figure~\ref{fig:simple_numlin_j}. The general relativistic case and its equivalent, given for $d=3/14$, are indicated by the dashed lines in the green region.  
The range of the phase space curvature values $\mathcal{R}$ are constrained by Eq.~\eqref{eq:constraint}, and we consider only real model parameters, bounded by $|d|\leq a/\sqrt{3}$. 
\begin{figure}[ht!]
     \includegraphics[scale=0.33]{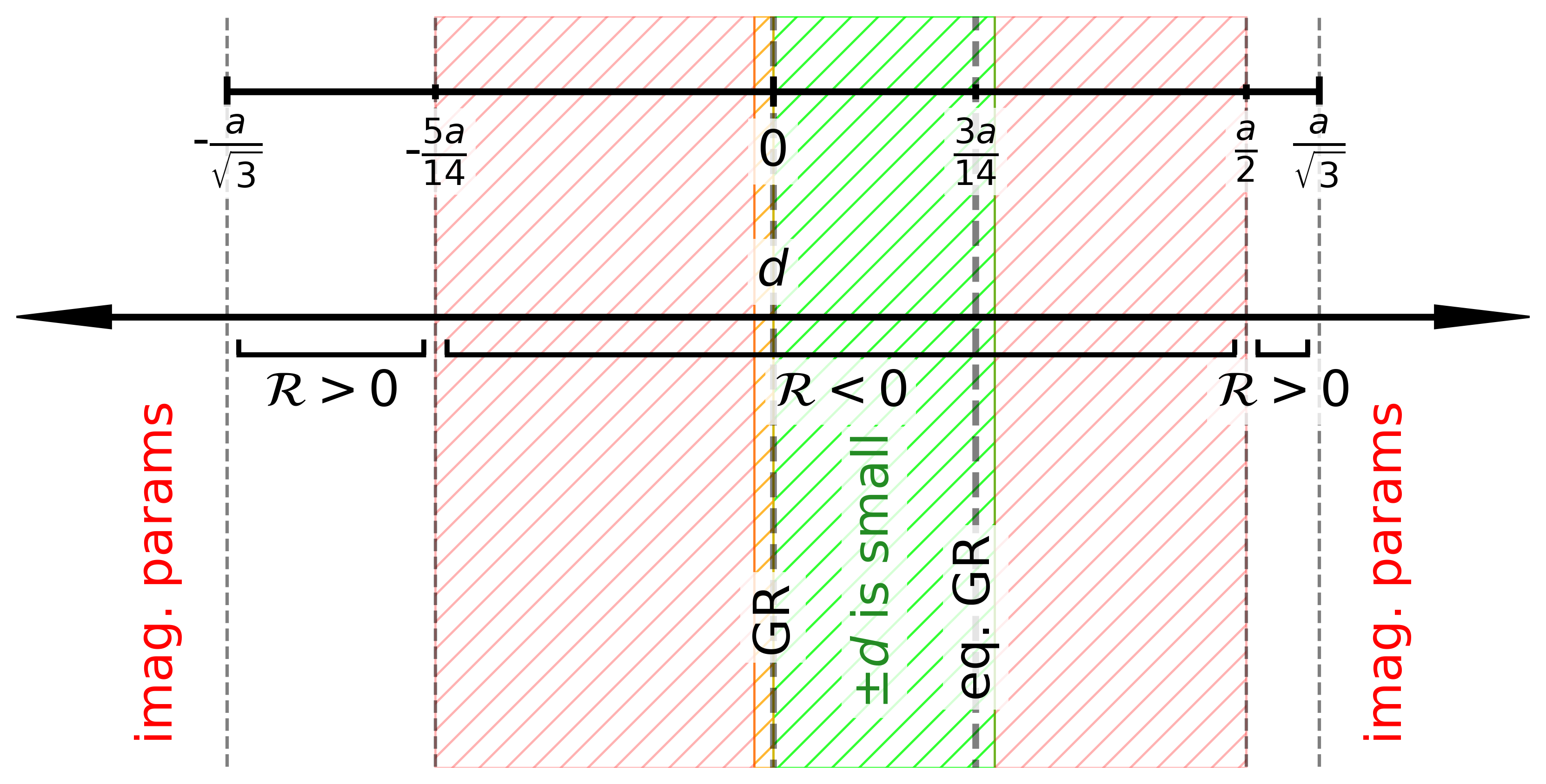}
     \caption{Properties of the Kaluza\,--\,Klein theory in the Jordan frame, shown for the parameter $d$, which is related to the scalar field, $\sigma$. Green region shows where the model is closest to general relativity, $b$ is set to be positive.}
     \label{fig:simple_numlin_j}
\end{figure}

Focusing on the phase space curvature of the Kaluza\,--\,Klein theory in the Jordan frame, Fig.~\ref{fig:simple_numlin_j} shows the different regions of the parameter space, based on Fig.~\ref{fig:ph_ratio2}. The most interesting part is included in the green region, where $d$ is small:
$$d\in\left[0, \frac{3a}{14}\right]$$
and which is the closest approximation to GR. Outside of this, however, one can also consider more exotic cases, see the discussion in Appendix~\ref{cases}. In Appendix~\ref{appendix_new1} special values of the phase space curvature are shown on the full parameter space. From now on, we are ready to analyse the physical consequences of the properties of the KK phase space curvature.

\section{Modified dispersion relation}
\label{sec:disp}

Let us briefly recall the expectation values of the momenta~\eqref{momentum}. For $\hat{N}^i=0$, the canonical and the physical momenta become equal, and the square of the momentum operator is given by~\cite{petruzziello2021gravitationally} 
\begin{equation}
    \hat{p}^2 = \hat{\pi}^2 ,
\end{equation}
where $\hat{\pi}^2\psi=-\hbar^2 \Delta \psi$, with $\Delta$ being the Laplace\,--\,Beltrami operator, corresponding to the square of the momentum operator, evaluated with respect to $G_{ij}$. Its uncertainty $\sigma_\mathrm{p}=\sqrt{\langle \hat{p}^2\rangle-\langle \hat{p}_i\rangle\langle \hat{p}^i\rangle}$ perturbed around flat spacetime ($a/r \ll 1$) reads
\begin{equation}
    \sigma_p \simeq \sqrt{  (\sigma_p^2) ^{(0)} +(\sigma_p^2) ^{(1)} + (\sigma_p^2)  ^{(2)} }\ .
\end{equation}
The zeroth and second order terms are non-zero, but the first order one disappears, $(\sigma_p^2) ^{(1)}=0$. Evaluating them for the ground state, $|\psi_{100}\rangle$ provides \cite{petruzziello2021gravitationally,dabrowski2020asymptotic}
\begin{align}
    (\sigma_p^2)^{(0)} \geq& -\hbar^2 \langle\psi_{100}|\Delta \psi_{100}\rangle^{(0)} = \frac{\hbar^2\pi^2}{\rho^2}\ ,\\
    (\sigma_p^2)^{(2)} =& -\frac{1}{6}\hbar^2 \mathcal R|_\mathrm{x_0}\ ,
\end{align}
leading to the modified uncertainty relation~\eqref{finalunc}.

Now, since we are in the co-moving frame, the expectation value squared $\langle\hat{p}\rangle^2$ vanishes, so one can write $(\sigma_p^2)\simeq \langle\hat{p}^2\rangle$. On the other hand, the 3-momentum $p_i$ specified by Eq.~\eqref{momentum} can be treated as the spatial part of the 4-momentum $p_\mu=(p_0,p_i)$ in curved spacetime. Considering the Jordan frame metric~\eqref{metric} as the background for a test particle with mass $m$, the square of the four-momentum for a co-moving observer can be written as 
\begin{equation}\label{psqr}
    p^\mu p_\mu = g_{00}(p^0)^2 + \hbar^2\left(\frac{\pi^2}{\rho^2} - \frac{\mathcal R}{6}\right)\ .
\end{equation}
Denoting the unperturbed 4-momentum in flat spacetime by $k_\mu=(k_0,k_i)$, with $p_0=k_0$, for which 
\begin{equation}
    k^\mu k_\mu = g_{00}(k^0)^2 + \mathbf{k}^2 = -m^2 \ ,
\end{equation}
is the usual dispersion relation, one can write
\begin{equation}\label{psqr2}
    p^\mu p_\mu = k^\mu k_\mu  - \frac{\hbar^2 \mathcal R}{6} = -m^2  - \frac{\hbar^2 \mathcal R}{6}
\end{equation}
and identify $\hbar^2\pi^2/\rho^2=\mathbf{k}^2$. 
Since the energy of a particle in a gravitational background is $E = - g_{00} p^0$, we can also express Eq.~\eqref{psqr} as
\begin{equation}\label{disp}
   E^2 = - g_{00} \left[m^2 + \left(\mathbf{p}^2 +  \frac{\hbar^2\mathcal R}{6}\right)\right]\ .
\end{equation}
The quantity 
\begin{equation}\label{effective}
    m_\text{eff} = \sqrt{m^2 +\frac{\hbar^2\mathcal R}{6}}
\end{equation}
can then be interpreted as the effective mass of a particle with mass $m$ in the gravitational field of the Schwarzschild-like black hole\footnote{Note that this effective mass expression is also valid for the GR case.}.

\begin{figure}[ht]
     \includegraphics[scale=0.55]{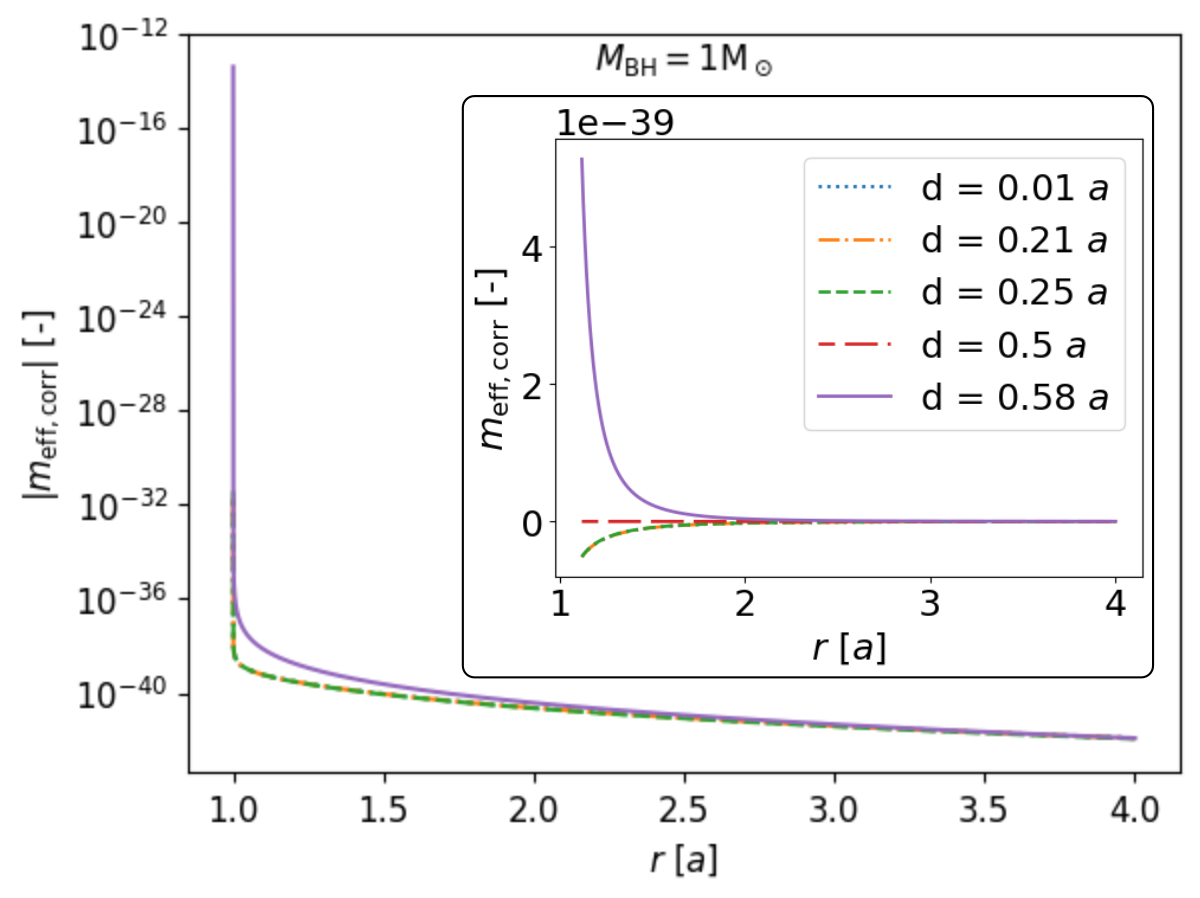}
     \caption{Absolute value of the approximate mass correction normalized to the neutron mass, in the Jordan frame. The calculation is specific to a black hole with mass $M_\mathrm{BH}=1$M$_\odot$, in SI units. A zoomed in version with sign on a linear scale is shown in the small plot.}
     \label{fig:masscorr}
\end{figure}
In Figure~\ref{fig:masscorr}, we plotted the mass correction 
\begin{equation}
    m_\mathrm{eff,corr} \approx \frac{1}{2}\frac{\hbar^2c^2}{(2GM)^2}\frac{\mathcal R}{6m^2} 
\end{equation}
 in the units of SI, normalized by the particle mass squared: here, $m=m_\mathrm{n^0}$ denotes the mass of neutron, while we also specify that $M_\mathrm{BH}=1$M$_\odot$. The radial coordinate was left in units of $a$ for clarity, with $r[a]=r[\mathrm{m}]\frac{c^2}{2GM}$, where $r[\mathrm{m}]$ is the radial distance provided in meters. Note that the main figure has a logarithmic scale on the vertical axis and shows the absolute value, while the small zoomed in version has a linear scale and shows the sign. Also, the special values of $d$, $3/14a\approx 0.21a$ and $a/\sqrt{3}$ is slightly larger than $0.58a$. The GR case, $d=0$ is indistinguishable from the technically calculable $d=0.01a$. 

 Far away from the central object, the modification of the effective mass is negligible as expected. However, at the horizon the correction diverges. It means that the effective mass can become imaginary, since the phase space curvature for the most interesting cases, the general relativistic and the Kaluza\,--\,Klein with $-5/14<d<1/2$ is negative. To assess the strength of this effect, one can examine the length scale at which it becomes significant.

In Horváth et. al.~\cite{Horvath:2024raf} mass corrections arise because of particle excitations in a compactified extra dimension. There the effective mass  (in SI units) reads: 
\begin{equation}\label{eq:meff_old}
    m^2_\mathrm{eff}=m^2+\hbar^2\frac{ N_\mathrm{exc}^2 }{c^2r^2_\mathrm{c}} \ ,
\end{equation}
where $N_\mathrm{exc}$ is the excitation number, $r_\mathrm{c}$ is the size of the extra dimension.  
Corrections were on the order of a few hundred MeV, which corresponds to an extra dimension size of about 0.2\,--\,1.0~fm. These corrections can already be present in the interior of neutron stars, that is, in a gravitationally less energetic environment than black holes. In the current work, the exact expression for the effective mass squared (in SI units) can be written as 
\begin{equation}\label{eq:meff_new}
    m^2_\mathrm{eff}=m^2+ \frac{c^2\hbar^2}{4G^2M^2}\frac{\mathcal R}{6} \ . 
\end{equation}
To get an effective mass correction for a 1M$_\odot$ black hole on the same order (regardless of sign) as in Ref.~\cite{Horvath:2024raf}, such that Eqs.~\eqref{eq:meff_old}~and~\eqref{eq:meff_new} are numerically comparable, the phase space curvature would need to be extremely large. To match for example the mass of the $\Lambda^0$ baryon, such that $m_\mathrm{eff}=m_{\Lambda^0}$, with the neutron as ground state, $m=m_\mathrm{n^0}$ and $r_\mathrm{c}=0.33$~fm, $\mathcal{R}$ needs to be on the order of $10^{38}$m$^{-2}$. Thus we can see in Fig.~\ref{fig:masscorr} that in order to reach a similar effect as described in Ref.~\cite{Horvath:2024raf}, the particle needs to be extremely close to the horizon of the black hole, where the phase space curvature diverges.   

\section{Conclusion}\label{sec:conc}

In this manuscript, we analyze the geometric properties of the Kaluza\,--\,Klein empty spacetime and its associated curved phase space. These results are compared to the general relativistic case both in the strong- and the weak-field limit. In the latter, the Kaluza\,--\,Klein Schwarzschild-like solution acquires additional terms due to the influence of the fifth dimension. We also show that the curvature of phase space modifies the uncertainty relation and the dispersion relation of massive particles, resulting in effective, geometry-dependent test particle masses.

The static, spherically symmetric solution~\cite{coquereaux1990theory} we start our considerations with is provided in the Einstein conformal frame, in which the field equations take a simple form. However, we consider the Jordan frame as the physical one, so our analysis focuses mostly on that.

First, we calculated the spacetime curvature $R$. The normalized expressions in the weak-field limit are presented in Table~\ref{table:pici}. It vanishes when considered in the Jordan frame, as for GR. However, in the Einstein frame it takes a non-zero value due to the contribution of the scalar field, and gets a correction quadratic in $d/a$. Parameter $d$ is proportional to the scalar field induced by the fifth dimension, while $a$ is analogous to the mass parameter in the Schwarzschild solution. Thus, based on the spacetime curvature, the behaviour of the Kaluza\,--\,Klein theory coincides with that of GR when the ratio $d/a \ll1$. This is in accordance with the calculations performed in Section~\ref{sec:adm}, regarding the Jordan frame ADM mass~\eqref{adm_mass}. Note that the existence of a non-zero scalar field-dependent mass correction at spatial infinity can have cosmological implications. For example, in certain models compact astronomical objects, or even galaxies in the weak-field limit could couple to the expansion of the universe and act as dark energy while gaining mass related to the scale factor~\cite{Farrah_2023,Cadoni:2024jxy,delima2025schwarzschilddesitterspacetimeregular}.

\begin{table}[]
\renewcommand{\arraystretch}{2.5}
\begin{tabular}{lcr}
\hline
\textbf{Curvature limits} ($b>0$) & $R/a^2$                         & $\mathcal{R}/a^2$                                                          \\ \hline \hline
General Relativity (GR)                                                 & 0                               & $\left[-\frac{5}{2}\right] \frac{1}{4 r^4}$                                                          \\ 
Kaluza\,--\,Klein (JF)                                              & 0                               & $ \left[ -\frac{5}{2}-3\frac{d}{a}+13\frac{d^2}{a^2} \right] \frac{1}{4 r^4}$ \\ 
Kaluza\,--\,Klein (EF)                                              & $\left[ 6\frac{d^2}{a^2}\right]  \frac{1}{4 r^4} $ & $ \left[ -\frac{5}{2}+\frac{27}{2}\frac{d^2}{a^2}\right] \frac{1}{4 r^4}$                           \\ \hline
\end{tabular}
\caption{Weak field limit ($r\to\infty$ up to the fourth order and $d\to 0$ up to the second order using power series expansion) of the normalized spacetime ($R$) and phase space ($\mathcal{R}$) Ricci scalar curvatures for GR and the KK theory both in the Jordan (JF) and in the Einstein (EF) conformal frames.}
\label{table:pici}
\end{table}

In the following, we have recalled the Hamiltonian formalism to analyze the motion of a massive particle in a curved background provided by the Kaluza\,--\,Klein Schwarzschild-like solution. Based on the quantum mechanical treatment of Petruzziello et.al.~\cite{petruzziello2021gravitationally}, one obtains that the uncertainty and dispersion relations of a massive particle are modified by a term proportional to the phase space curvature $\mathcal{R}$. It is present in the Kaluza\,--\,Klein model in both frames, and even in GR with different limiting behaviours, see Table~\ref{table:pici}. It naturally leads to a modified effective mass of massive particles in both theories, see Section~\ref{sec:disp}.

Is there an opportunity to observe this effect in the strong-field regime? One of the possible tests is through establishing a modified thermodynamics of particles~\cite{PhysRevD.107.044025}, which could lead to modified white dwarf and neutron star mass-radius relations, as well as other observables, e.g. luminosity, or surface temperature. Previously, we have demonstrated how particles moving in extra dimensions have an effect on the maximal mass of neutron stars through the equation of state~\cite{Horvath:2024raf,Horvath:2024qhk}. Another interesting aspect to consider is a modified speed of sound~\cite{Horvath:2025sdg}, which can have a signature on dynamical phenomena, e.g. gravitational wave signals.   

Note that for certain values of the scalar field parameter $d$, the phase space curvature vanishes in the full relativistic case. Then one deals with no modification in either the uncertainty principle or the effective mass. Notice that it does not have to result in an undeformed Heisenberg algebra \cite{Pachol:2023tqa,Bosso:2023aht}, however, it would be rather impossible to test such subcases.

Significant effects can be expected near black hole horizons. In GR and in the $d\in\left(-\frac{5a}{14}, \frac{a}{2}\right)$ regime of the Kaluza\,--\,Klein theory the phase space curvature is negative, introducing a positive correction to the uncertainty relation. At the same time, however, the effective mass can become imaginary, as the phase space curvature diverges. One may interpret such an effect as a particle decay, for its time evolution is now described by an exponential which has a real negative exponent $\sim\ee^{iEt}\approx \ee^{imt}$ instead of just a phase~\cite{PhysRevD.99.065017}. This decay induced by a strong gravitational field could be associated for example with Hawking radiation~\cite{Falcke_2025}, or it could affect the energy loss of supernovae~\cite{HANHART2001335}.

Another possibility is to consider values of the scalar field parameter from $d\in\left[-\frac{a}{\sqrt{3}},-\frac{5a}{14}\right) \cup \left(\frac{a}{2},\frac{a}{\sqrt{3}}\right]$, where $\mathcal{R}>0$. In this case, although the effective mass will always be positive, the uncertainty relation could constrain the parameter $d$ if $\mathcal{R}$ gets large enough. Recall that we do not take into account the part related to the shift vector in the modified uncertainty relation. Also, notice that the value of the phase space curvature diverges at the horizon, possibly breaking the uncertainty relation in this simplified approach. 

In this framework, it is also possible to consider imaginary model parameters, that is, one can deal with a complex spacetime structure. However, this can happen in different ways, and one of the major cases is when the scalar field parameter $d$ is real (so the scalar field, too) but $b$ appearing in the solution \eqref{metric} becomes complex. On the other hand, we see that the parameter $d$ can also become complex. Therefore, one deals with a complex scalar field, which as mentioned before, is currently a popular model of dark matter, for example by modeling boson stars. Nevertheless, the specific case of complex model parameters and its potential physical implications are left for future investigation.

In general, based on the results of this work, the observable consequences of an extra dimension or the presence of a scalar field can constrain the model parameters. As long as one keeps the $d/a$ ratio sufficiently low, it leaves room for the existence of new physics. With larger $d/a$, or imaginary model parameters more exotic solutions are also possible.

\section*{Acknowledgements}

AH would like to acknowledge the support given by the HUN-REN KMP program under contract numbers KMP-2023/101 KMP-2024/31 and KMP-2025-I/45, the ERASMUS+ Short-term mobility doctoral program, the BridgeQG (CA23130) COST Action's Short Term Scientific Missions program for in person collaboration opportunities, and the DKÖP program of the Doctoral School of Physics at the Eötvös Loránd University.  

AW acknowledges financial support from MICINN (Spain) {\it Ayuda Juan de la Cierva - incorporaci\'on} 2020 No. IJC2020-044751-I and from  the Spanish Agencia Estatal de Investigaci\'on Grant No. PID2022-138607NB-I00, funded by MCIN/AEI/10.13039/501100011033, EU and ERDF A way of making Europe.

GGB acknowledges K135515, 2024-1.2.5-TÉT-2024-00022, 2021-4.1.2-NEMZ\_KI-2024-00031, 2025-1.1.5-NEMZ\_KI-2025-00005,.

The authors would like to express their gratitude to Anna Pachoł, Fabian Wagner, Christian Pfeifer, Dániel Barta and Pok Man Lo for their inspiring discussions. They also acknowledge networking support by the COST Action FuSe CA24101.

AW would like to express her gratitude to the members of the HUN-REN Wigner Research Centre for Physics, Budapest, Hungary, for their warm hospitality during her stay while working on this manuscript.
\bibliographystyle{apsrev4-1}
\bibliography{biblio}

\appendix

\section{Parameter space}
\label{appendix_new1}

\begin{figure}[ht]
     \centering     
     \includegraphics[scale=0.55]{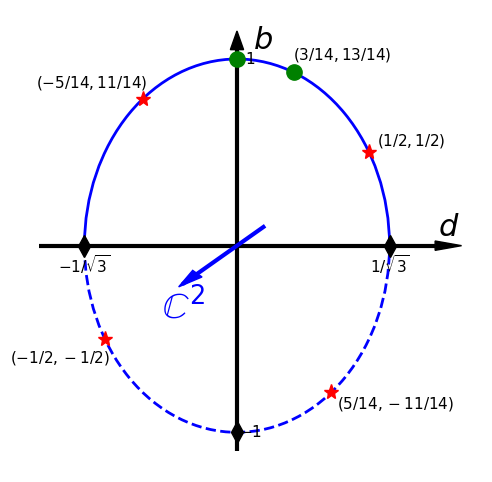}
     \caption{Parameter space spanned by $b$ and $d$ with $a=1$ in the Jordan frame. Numbers are shown in units of $a$. Special values are indicated by markers.}
     \label{fig:param_space}
\end{figure}

The whole parameter space of the Kaluza\,--\,Klein Schwarzschild-like solution in the Jordan conformal frame (spanned by $b$ and $d$ with $a=1$) is presented in Figure~\ref{fig:param_space}. Possible solutions are located on an ellipse in the real plane fixed by constraint~\eqref{eq:constraint}. If we allow imaginary solutions, values are confined to a 4-dimensional hypersurface with either $b$ or $d$, or both being complex. The special values in the real plane marked by red stars correspond to zeros of the expression
\begin{equation}
    4a^2-(3b+d)^2 :=0\ ,
\end{equation}
coming from the phase space curvature~\eqref{curv2}. At these values $\mathcal{R}$ changes sign. The green dot markers correspond to the trivial $d=0$ GR case, and an equivalence to GR, where Eq.~\eqref{curv2} takes an identical form. Black diamonds show the extrema of the parameter space with either $b$ or $d$ being zero, separating different regimes. The $b>0$ part, examined in the other figures is drawn with a solid curve, while the $b<0$ region is dashed. 

\section{Kaluza\,--\,Klein and other theories}
\label{cases}

Let us consider all of the cases following from Fig.~\ref{fig:simple_numlin_j}, where we set $b\in[0,a]$: 

\begin{description}
\item[Close approximation to GR ] $$d\in\left[0, \frac{3a}{14}\right]$$ 
This is the most interesting region in which $d$ has a small and positive value, not differing too much from the GR case (green area). Moreover, it can be subject to observational constraints. Trivially, $d=0$ recovers GR, while the $d=\frac{3a}{14}$ case is particularly special, since although the phase space curvature is the same as in GR everywhere, we have a non-zero scalar field present. With this, a case in which an alternative theory gives the same predictions as GR is possible -- the value of $\mathcal{R}$ modifying the uncertainty relation coincides. See further discussion in Sec.~\ref{sec:disp}.

\item[Negative phase space curvature ($\mathcal{R}<0$)] $$d\in\left(-\frac{5a}{14}, \frac{a}{2}\right)$$ The next broader region to consider in Fig.~\ref{fig:simple_numlin_j} is  
where the phase space curvature is negative, similarly to GR (red and green areas). However, the deviance from Einstein's relativity is quite significant towards the edges of the regime, and the $d<0$ regime corresponds to a negative scalar field for the considered $a>0$. 

\item[Vanishing phase space curvature ($\mathcal{R}=0$)] $ $ \\[2ex]
At the points $d=-\frac{5a}{14}$ and $d=\frac{a}{2}$ there is no modification to the uncertainty relation.

\item[Positive phase space curvature ($\mathcal{R}>0$)]
$$d\in\left[-\frac{a}{\sqrt{3}},-\frac{5a}{14}\right) \cup \left(\frac{a}{2},\frac{a}{\sqrt{3}}\right]$$
In these regimes 
the phase space curvature becomes positive. It provides a modification to the uncertainty relation~\eqref{finalunc} allowing, in principle, for a decrease in the uncertainty or even making it zero. Moreover, the phase space curvature diverges at the horizon, see also the discussion in Sec.~\ref{sec:conc}. Note that $d=\pm\frac{a}{\sqrt{3}}$ provides $b=0$.

\item[Imaginary model parameters] Outside of the regions discussed so far, at least one of the model parameters ($b$ or $d$, with $a=1$) must become imaginary by Eq.~\eqref{eq:constraint}.
\begin{itemize} 
     \item $b\in\mathbb{C}$, 
 $d\in\left(-\infty,-\frac{a}{\sqrt{3}}\right)\cup\left(\frac{a}{\sqrt{3}},+\infty\right)$

For these values one deals with solutions being far away from the Schwarzschild one. Since one deals with a complex metric (because of the imaginary $b$), one can consider a complex spacetime geometry, in a similar manner as in the case of the complex scalar field (see the comment below). We do not analyse this case in this paper.

\item $d\in \mathbb{C}$, $b>1a$

For values of $b$ from this region,
the scalar field parameter becomes imaginary. Note that the idea of a complex scalar field in 4-dimensional real spacetime introduces exotic compact objects, such as boson stars \cite{colpi1986boson} and/or some models of dark matter. On the other hand, in the 5-dimensional case the existence of a complex scalar field means that we are dealing with a complex metric~\eqref{5metric}, see e.g. \cite{lehners2023allowable}, providing a complex spacetime geometry, in which the real part corresponds to mass and gravitation, while the imaginary part to charge and electromagnetism \cite{trautman1962analytic,newman1973maxwell,einstein1946generalization,plebanski1975some}. Although the complex scalar field has gained a lot of attention recently, we will not focus on it in this work.
\end{itemize}
\end{description}

\end{document}